\documentclass[useAMS,usenatbib]{mn2e}

\usepackage{psfig, epsf, epsfig,float}

\title[The origin of BCDs]{Formation and evolution of blue compact dwarfs:
the origin of their steep rotation curves}
\author[A. Watts and K. Bekki]
{Adam  Watts${}^1$\thanks{E-mail:
21146752@student.uwa.edu.au} 
and
Kenji Bekki${}^1$\thanks{E-mail: kenji.bekki@uwa.edu.au} \\
${}^1$ICRAR M468
The University of Western Australia
35 Stirling Hwy, Crawley
WA 6009, Australia}

\begin{document}

\date{Received 2016 March 23; in original form. Revised 2016 July 8. Accepted for publication in MNRAS on 2016 July 21. }

\pagerange{\pageref{firstpage}--\pageref{lastpage}} \pubyear{2016}

\maketitle

\label{firstpage}

\begin{abstract}

The  origin of the observed
steep rotation curves of blue compact dwarf galaxies (BCDs) remains 
largely unexplained by theoretical models of BCD formation.
We therefore investigate the rotation curves in BCDs formed from mergers 
between gas-rich dwarf irregular galaxies
based on the results of numerical simulations for BCD formation.
The principal results  are as follows.
The dark matter of merging dwarf irregulars undergoes 
a central concentration so that the central density can become 
up to six times higher than 
those of the initial dwarf irregulars. 
However,  the more compact dark matter halo
alone can not reproduce the gradient differences observed 
between dwarf irregulars and BCDs.
 We provide further support that the
central concentration of  gas due to rapid gas-transfer to the
central regions of dwarf--dwarf mergers
is responsible for the observed 
difference in rotation curve gradients. 
The BCDs with central gas concentration
formed from merging can thus show steeply rising rotation
curves in their central regions.
Such  gas concentration is  also responsible
for central starbursts of BCDs and  the high central surface brightness  and is consistent with previous BCD studies. 
We discuss the relationship between rotational velocity gradient and surface brightness, the dependence of BCD rotation curves on star formation threshold density, progenitor initial profile, interaction type and merger mass ratio, as well as potential evolutionary links between dwarf irregulars, BCDs and compact dwarf irregulars.
\end{abstract}


\section{Introduction}
Blue compact dwarf galaxies (BCDs) resemble the spectra of H{\sc ii} regions of spiral galaxies while being optically small, low luminosity (Thuan \& Martin 1981),  and typically exhibit subsolar metallicities  [Fe/H] $< -1$ (Drozdovsky \& Tikhonov 2000; Drozdovsky et al. 2001). Many studies have turned their focus to BCDs as they exhibit features similar to those of high-redshift galaxies  (Papaderos 2006) and were originally thought to have been undergoing their first starbursts (e.g. Searle, Sargent \& Bagnuolo 1973; Aloisi, Tosi \& Greggio 1999; Thuan, Izotov \& Foltz 1999), making them potential insights into the early Universe. The more recent confirmation of an underlying population of old stars in most BCDs (e.g. James 1994; Cair{\'o}s et al. 2001; Amor\'in et al. 2007) did not compromise their potential as valuable low-redshift laboratories to study star formation and galaxies analogous to the early Universe due to their low metallicities (Thuan et al. 1995; Izotov \& Thuan 1999). The rotation curves of dwarf galaxies generally resemble solid body rotation in their central regions, but when compared to their disc scale length they rise as steeply as spirals (Swaters et al. 2009). Comparatively, BCDs exhibit central rotation curves significantly steeper than the solid body rotation profile implying a strong central concentration of mass (Lelli et al. 2012a,b; Lelli et al. 2014a, hereafter L14; Koleva et al. 2014, hereafter K14). Dark matter of BCDs is observed to have high central densities around 0.1 M$_{ \odot} $ pc$^{-3}$ (Meurer et al. 1996; Meurer, Stavley-Smith \& Killeen 1998, hereafter M98) and centrally peaking H{\sc i} maps  (e.g. van Zee et al. 1998; Lelli et al. 2014b) have frequently been related to their elevated central starburst regions and small optical sizes (e.g. McQuinn et al. 2015).

One of the unresolved problems in BCDs
is their steep central rotational velocity profiles
(e.g. M98; Lelli et al. 2012a,b; L14).
Mass models of the BCDs NGC 2915 and NGC 1705 show rotation curves dominated by dark matter with central dark matter densities to up to 10 times higher than those of dwarf irregulars (Meurer et al. 1996; M98; Elson, de Blo \& Kraan-Korteweg 2010) but the definite cause of this concentration remains undecided. With it becoming more obvious that the concentration of not only dark matter is necessary for BCD formation but the concentration of  gas also,
several mechanisms 
have already been suggested. Torques from internal gas and star clumps (Elmegreen et al. 2012) and rotating triaxial dark matter haloes (Bekki \& Freeman  2002) are among the many proposed.   Verbeke et al. (2014, hereafter V14) showed that in-falling gas clumps consistently increase central concentration of gas  consequentially causing a steeper rotation curve, ignites starbursts extending over a few 100 Myr and increases central surface brightness in simulated dwarf galaxies.
Another mechanism that has been gaining rapid support recently is dwarf--dwarf galaxy mergers (e.g. Noeske et al. 2001; {\"O}stlin et al. 2001; Bekki 2008, hereafter B08). Recent observational studies frequently find that gas distributions in BCDs resemble that of merger remnants (e.g. Ekta et al. 2008), the majority of BCDs show signs of tidal interaction or galaxy merging in their starburst component (e.g.  Pustilnik et al. 2001) and Pak et al. 2016 observed an apparent in-progress merger with two blue star forming cores and exponential surface brightness.  However, even though dwarf galaxy mergers efficiently transport large amounts of gas to the centre of the interaction 
(Bekki 2015, hereafter B15),
these  models have not yet discussed
whether BCDs formed from dwarf--dwarf merging 
explain the observed steep rotation curves of BCDs (L14).

The purpose of this paper is to investigate the rotational velocity profile of BCDs  produced through merger interactions
based on a set of new hydrodynamical simulations with star formation,
chemical evolution, and dust evolution.
A  primary focus of this paper 
is on the origin and evolution of steep rotation curves
of BCDs.
We mainly investigate 20 models of BCDs formed from dwarf--dwarf
merging with different star formation threshold densities,
merger mass ratios, and merger orbits.
The rotation curves of galaxies are a 
useful tool for investigating mass distributions of different
components (e.g. stellar disc and dark matter halo) in galaxies
and distinguishing between contributions from each of these to the
mass distributions.
Therefore, the present numerical study contributes greatly to the better
understanding of the origin of the observed mass profiles of BCDs.
The present study is complementary to previous simulations of BCD formation
and evolution  through dwarf--dwarf galaxy mergers (e.g. B08; B15) 
because the major focuses of
these previous studies were mostly on structures,
metallicities, and post-merger evolution of BCDs.

The plan of the paper is as follows. First,
we describe the details of the adopted dwarf--dwarf merger model of
BCD formation.
Then, we present some key results of the new simulations on the 
rotation curves of BCDs formed from dwarf--dwarf mergers
and their dependence on the model parameters of dwarf--dwarf merging.
In this section, we also discuss how the steep rotation curves of BCDs
can be achieved during dwarf--dwarf merging.
In \S 4, we discuss the importance of major merging, star formation
threshold gas density, and initial dark matter mass profiles
in the formation of the steep rotation curves of BCDs. 
We summarize our  conclusions in \S 5.

\section{The model}
\subsection{Dwarf irregular galaxies}

We investigate the dynamical properties of the remnants
of  mergers between dwarf irregular
galaxies in order to understand the origin of BCDs. 
In order to simulate the time evolution of
rotation curves, star formation rate (SFR), and  gas and stellar  contents of merging dwarfs,
we use our original chemodynamical simulation code that can be run
on GPU machines (Bekki 2013, hereafter B13, 2014, hereafter B14).
The  code combines the method of smoothed particle
hydrodynamics (SPH) with \textsc{grape} for calculations of three-dimensional
self-gravitating fluids in astrophysics so that dynamical properties of gas and stars
in mergers can be self-consistently investigated.
Our main interest is the time evolution of dynamical properties
and SFRs in dwarf mergers (not the evolution
of dust and ${\rm H}_2$ contents). Therefore,
we do not incorporate ${\rm H}_2$ calculations (that can be done with the code)
into the present simulations.

A dwarf galaxy is composed of  dark matter halo,
stellar disc,   and  gaseous disc (and no bulge).
The total masses of dark matter halo, stellar disc, and gas disc
are denoted as $M_{\rm h}$, $M_{\rm s}$, and $M_{\rm g}$, respectively.
The mass ratio of gas to stars
and that of dark matter to stars
are denoted as $f_{\rm g}$ and $f_{\rm dm}$, respectively,  for convenience.
We mainly investigate the models with
$M_{\rm h}=10^{10} {\rm M}_{\odot}$, $f_{\rm dm}=167$,
and $f_{\rm g}=4$ to model low-mass, gas-rich dwarf irregular galaxies.

The density profile of the dark matter halo
of a dwarf irregular  is represented by that proposed by
Salucci \& Burkert (2000):
 
\begin{equation}
{\rho}(r)=\frac{\rho_{\rm 0,dm}}{(r+a_{\rm dm})(r^2+{a_{\rm dm}}^2)},
\end{equation}
where $\rho_{\rm 0,dm}$ and $a_{\rm dm}$ are the central dark matter
density multiplied by $a_{\rm dm}^3$
and the core (scale) radius, respectively.

For convenience, we hereafter call this profile the ``SB'' profile.
The core radius is set to be $0.1 r_{\rm vir}$, where $r_{\rm vir}$ is the virial
radius of the dark matter halo ($r_{\rm vir}=14.0$ kpc 
for $M_{\rm h}=10^{10} {\rm M}_{\odot}$).
The main difference between the SB profile and the Navarro--Frenk--White (NFW) profile
halo (Navarro, Frenk \& White 1996) suggested from cold dark matter (CDM) simulations
is that the SB  has a large dark matter core (i.e., low central mass density).
Following the recent results by Oh et al. (2011),
we choose the SB profile rather than the NFW profile for most of the models.

However we also investigate the models with the following NFW profile
for comparison in the present study:
\begin{equation}
{\rho}(r)=\frac{\rho_{0}}{(r/r_{\rm s})(1+r/r_{\rm s})^2},
\end{equation}
where  $r$, $\rho_{0}$, and $r_{\rm s}$ are
the spherical radius,  the characteristic  density of a dark halo,  and the
scale
length of the halo, respectively.
The $c$-parameter ($c=r_{\rm vir}/r_{\rm s}$, where $r_{\rm vir}$ is the virial
radius of a dark matter halo) and $r_{\rm vir}$ are chosen appropriately
for a given dark halo mass ($M_{\rm dm}$)
by using the $c-M_{\rm h}$ relation
predicted by recent cosmological simulations (Neto et al. 2007).
We adopt 
$c=16$ is an appropriate value because it is appropriate for $M_{\rm h}=10^{10} {\rm M}_{\odot}$.

The radial ($R$) and vertical ($Z$) density profiles of the stellar disc are
assumed to be proportional to $\exp (-R/R_{0}) $ with scale
length $R_{0} = 0.2R_{\rm s}$  (where $R_{\rm s}$ is the radius of the stellar disc) and to ${\rm sech}^2 (Z/Z_{0})$ with scale
length $Z_{0} = 0.04R_{\rm s}$, respectively.
The gas disc with a size  $R_{\rm g}=5R_{\rm s}$
has the  radial and vertical scale lengths
of $0.2R_{\rm g}$ and $0.02R_{\rm g}$, respectively.
The exponential disc
has $R_{\rm s}=1.75$ kpc and
$R_{\rm g}=8.75$ kpc for most models in the present study.
In addition to the
rotational velocity caused by the gravitational field of disc,
bulge, and dark halo components, the initial radial and azimuthal
velocity dispersions are assigned to the disc component according to
the epicyclic theory with Toomre's parameter $Q$ = 1.5.
The vertical velocity dispersion at a given radius is set to be 0.5
times as large as the radial velocity dispersion at that point.

%%%%% TABLE1
\begin{table*}
\centering
\begin{minipage}{160mm}
\caption{Description of the basic parameter values
for the isolated and merging  dwarf  models.
%The comparative models M7 and M8
%are those with no stars ($M_{\rm s}=0$; ``starless dwarfs'').
 }
\begin{tabular}{cccccccccc}
Model ID
& DM type
& $f_{\rm g}$
& $m_2$
& $e_{\rm orb}$
& Merger configuration
& $\rho_{\rm th}$ (atom cm$^{-3}$) & Comments \\
M1 & SB & 4.0 & -- & -- & -- & 10.0 & Fiducial isolated\\
M2 & SB & 8.0 & -- & -- & -- & 1.0 & High $f_{\rm g}$ \\
M3 & SB & 12.0 & -- & -- & -- & 1.0 & Very high $f_{\rm g}$\\
M4 & SB & 4.0 & -- & -- & -- & 1.0 & $R_{\rm g}=3R_{\rm s}$ \\
M5 & SB & 12.0 & -- & -- & -- & 10.0 & \\
M6 & NFW & 12.0 & -- & -- & -- & 10.0 & Steep inner DM profile (NFW) \\
M7 & SB & 4.0 & 1.0 & 1.0 & P & 10.0 & Fiducial merger\\
M8 & SB & 4.0 & 1.0 & 1.0 & P & 1.0 & Lower SF threshold density\\
M9 & SB & 4.0 & 1.0 & 0.6 & P & 1.0 & $R_{\rm g}=3R_{\rm s}$ \\
M10 & SB & 4.0 & 1.0 & 1.0 & P & 10.0 & $R_{\rm g}=3R_{\rm s}$ \\
M11 & SB & 8.0 & 1.0 & 1.0 & P & 1.0 &  \\
M12 & SB & 8.0 & 1.0 & 1.0 & R & 1.0 & \\
M13 & SB & 12.0 & 1.0 & 1.0 & P & 1.0 & \\
M14 & SB & 12.0 & 1.0 & 1.0 & R & 1.0 & \\
M15 & SB & 8.0 & 1.0 & 1.0 & R & 10.0 & \\
M16 & SB & 12.0 & 1.0 & 1.0 & R & 10.0 & \\
M17 & SB & 4.0 & 0.3 & 1.0 & P & 10.0 & Unequal-mass merger ($R_{\rm g}=3R_{\rm s}$) \\
M18 & SB & 4.0 & 0.3 & 1.0 & P & 10.0 & Unequal-mass merger\\
M19 & NFW & 4.0 & 1.0 & 1.0 & P & 10.0 & NFW DM profile\\
M20 & SB & 4.0 & 0.1 & 0.6 & P & 10.0 & Minor merger\\
\end{tabular}
\end{minipage}
\end{table*}

\begin{figure*}
\psfig{file=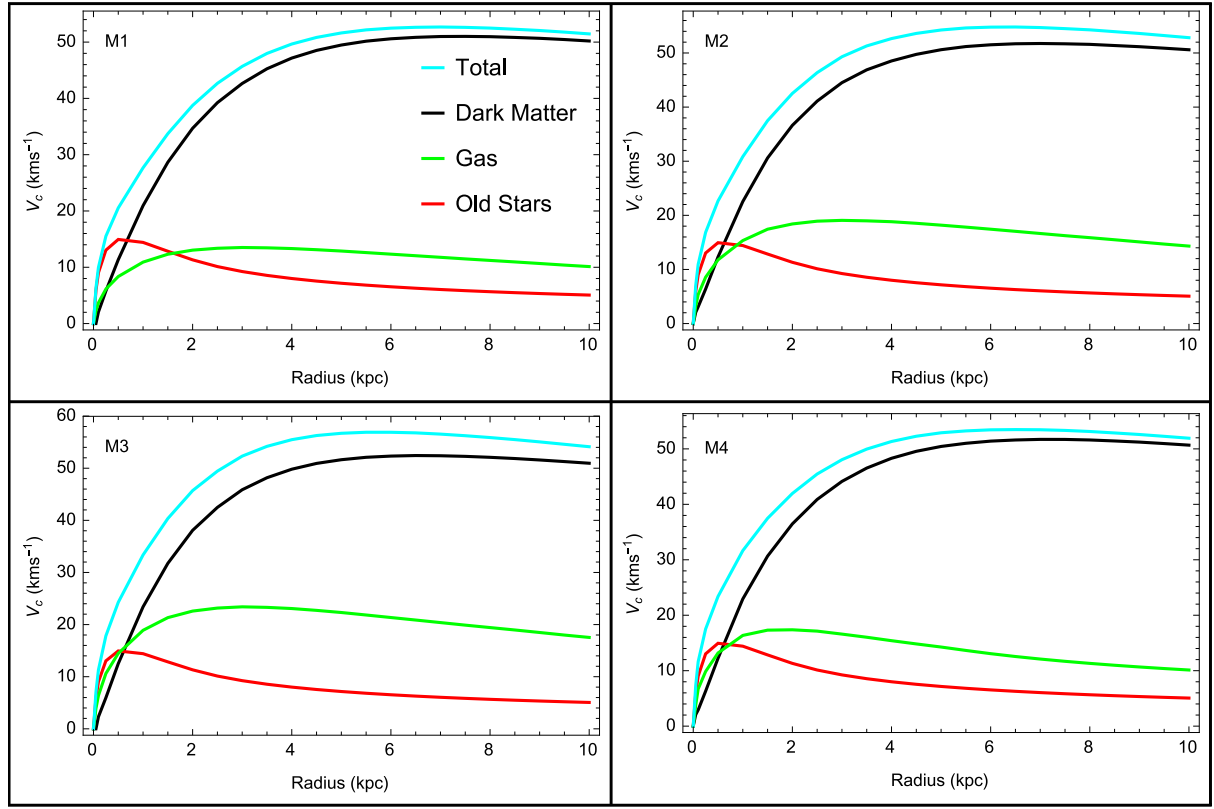,width=16cm}
\caption{The contribution of dark matter (black), gas (purple),  old stars (red), and  total (magenta) to the initial rotation curve of the isolated models M1 (top left), M2 (top right), M3 (bottom left), and M4 (bottom right). }
\label{Figure. 1}
\end{figure*}

\subsection{Star formation and SN feedback effects}

 The model for star formation in the present study
is essentially the same as those adopted in our previous simulations
(e.g. B13; B14). Therefore, we briefly describe it here.
A gas particle can be  converted
into a new star if (i) the local dynamical time-scale is shorter
than the sound crossing time-scale (mimicking
the Jeans instability) , (ii) the local velocity
field is identified as being consistent with gravitationally collapsing
(i.e., div  {\bf \textit{v}}$<0$),
and (iii) the local density exceeds a threshold density for star formation ($\rho_{\rm th}$).
Although B13 estimated the above three conditions using the local ${\rm H_2}$ density,
we here use local gas density (H{\sc i}+${\rm H_2}$) to investigate
whether the above three conditions are met.
The threshold gas density is given in units of the number of hydrogen atoms
per cm$^3$ in the present study just for convenience.
We mainly investigate the models with $\rho_{\rm th}=10$ cm$^{-3}$,
and the dependences of the present results on $\rho_{\rm th}$
are briefly discussed later.
The gas consumption rate due to star formation is determined by
the  Kennicutt-Schmidt law
(SFR$\propto \rho_{\rm g}^{\alpha_{\rm sf}}$;  Kennicutt 1998).
A reasonable value of
$\alpha_{\rm sf}=1.5$ is adopted in the present
study.

The model for supernova (SN) feedback effects is the same as those adopted in B13 and B14.
Each SN can  eject the feedback energy ($E_{\rm sn}$)
of $10^{51}$ erg and 90 and 10 percent of $E_{\rm sn}$ are used for the increase
of thermal energy (`thermal feedback')
and random motion (`kinematic feedback'), respectively.
The thermal energy is used for the `adiabatic expansion phase', where each SN can remain
adiabatic for $t_{\rm adi}=10^6$ yr in the present study.
The energy ratio of thermal to kinematic feedback is consistent with
previous numerical simulations by Thornton et al. (1998) who investigated
the energy conversion processes of SNe in  detail.
The way to distribute $E_{\rm sn}$ of SNe among neighbour gas particles
is the same as described in B13.
 The radiative cooling processes
are properly included  by using the cooling curve by
Rosen \& Bregman (1995) for  $100 \le T < 10^4$K
and the \textsc{mapping iii} code
for $T \ge 10^4$K
(Sutherland \& Dopita 1993).
The models for chemical evolution and dust growth are
the same as those adopted in B13.
The initial gaseous metallicity is set to be [Fe/H]=$-1.0$ dex
in all models.  Metals ejected from stars can be locked up either by interstellar medium (ISM; as gas-phase metal) or by
dust grains, which were  not considered in almost all simulations of galaxy formation and evolution.
Because of this,  the gas-phase metallicity is more properly estimated,  which can influence
metallicity-dependent radiative cooling (though it should be minor).

\subsection{Merger models}

Although we mainly investigate the rotation curve profiles of the remnants
of major dwarf--dwarf mergers, we also investigate
those of  isolated models for comparison.
We adopt the following merger models with parabolic ($e_{\rm orb}=1$,
where $e_{\rm orb}$ is the orbital eccentricity of a merger)
and elliptic orbits ($e_{\rm orb}=0.6$).
The initial distance
and the pericentre distance ($R_{\rm p}$)
of two interacting/merging dwarfs are set to be
$10R_{\rm s}$ and $0.5R_{\rm s}$, respectively, for the parabolic orbit model
whereas they are $6R_{\rm s}$ and  $2R_{\rm s}$, respectively, for the elliptic orbit
model.
The spin of each galaxy in an interacting or  merging pair  is specified by
angles $\theta_i$ (in units of degrees), where subscript $i$ is
used to identify each galaxy. $\theta_i$ is
the angle between the \textit{z}-axis and the vector of the angular momentum of a disc.
We show the results of the models with $\theta_1=30$ and $\theta_2=45$
(prograde merger; `P') and 
those with $\theta_1=150$ and $\theta_2=135$ (`retrograde merger; `R').
The mass ratio of two dwarfs is a free parameter denoted as $m_{2}$. We mainly
show the results of `major mergers' with $m_2=1$, because the simulated rotation 
curves are more consistent with the observed ones for BCDs.

Although we have run many isolated and merger models,
we mainly 
show the results of six isolated models (M1--M6) and 14 merger models
(M7--M20), because these models show important behaviours in the evolution
of rotation curves of dwarf galaxies.
We investigate the rotation curve profiles ($V_{\rm C}(r)$) for (i) BCD phases
when star formation is dramatically enhanced and (ii) remnant phases when merging
has been completed for each model.
We investigate the time evolution of velocity gradient (often noted $V_{\rm C}/R_{ \rm C}$ from here), which is defined as
the maximum circular velocity within a certain radius ($V_{\rm C}$)
divided by $R_{\rm C}$, for BCD and remnant phases in each merger model.
We also correlate the gradients ($V_{\rm C}/R_{\rm C}$) with
other global galaxy parameters such as surface or volume 
mass densities of dark matter,
gas, and stars.
The basic parameter values for the models are summarized in
Table 1.

\section{Results}
\subsection{Fiducial Model}
\subsubsection{Classification of BCDs}
Fig. 1 shows the rotation curves for four of the six  isolated models. Each major merger model consisted of a pair of dwarf galaxies with one of these six initial conditions and the top left  is the fiducial isolated model M1. The maximum observed circular velocity in these four models is around 60 km s$^{-1}$, the steepest gradient was approximately 55 km s$^{-1}$ kpc$^{-1}$, and the SFR for these models remained of order 10$^{-3}$ for the entire simulation.

Fig. 2 shows how two gas-rich dwarf galaxies evolve into one new compact dwarf with a high concentration of new (yellow) stars and outer extended  gas (cyan). As the two dwarfs interact, the distribution of old stars (magenta) becomes more diffuse and irregular whereas the  gas is efficiently attracted towards the centre of the interaction resulting in a dense envelope with very little net loss. This central concentration of  gas contains most of the interacting gas from the dwarf mergers and is clearly seen in Fig. 2 as an irregular envelope slightly smaller than the gas disc of the progenitors. This ignites an intense starburst resulting in the compact core of new stars that is rapidly produced and provides the high central surface brightness observed in BCDs.

\begin{figure}
\psfig{file=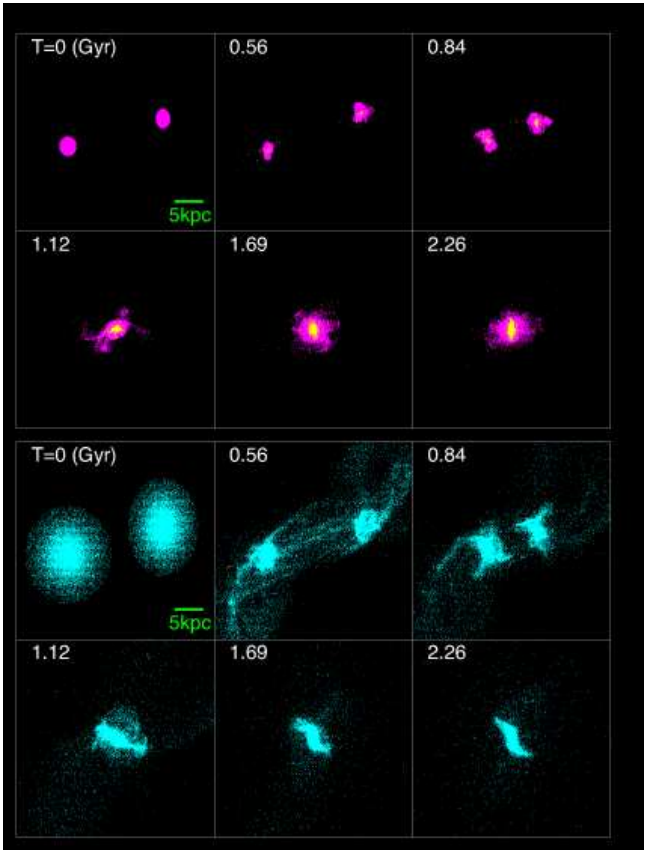,width=8.5cm}
\caption{The dynamics of the merger process for the fiducial model M7 showing gas (cyan), old stars (magenta), and new stars (yellow). One particle in every four of each component has been displayed. The warped and extended gaseous envelope of the BCD surrounds the central core of new stars while developing into what could resemble two small spiral arms.}
\label{Figure. 2}
\end{figure}

\begin{figure}
\psfig{file=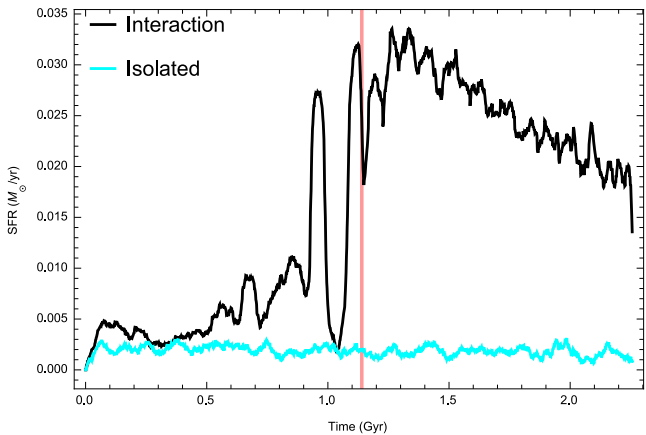,width=8.5cm}
\caption{Star formation histories of isolated model M1 and merger model M7. The red line indicates 1.12 Gyr  corresponding to the BCD phase. We observe the SFR to be increased by an order of magnitude as the BCD phase is entered.}
\label{Figure. 3}
\end{figure}

Fig. 3 further reflects this starburst as the SFR increases by an order of magnitude over the isolated model and remains elevated for the time-scale of a Gyr. The red line indicates a simulation time of 1.12 Gyr and marks the potential initiation of a BCD phase.  The maximum rotational velocity at this time is 62 km s$^{-1}$ compared to an initial value of 40 km s$^{-1}$ and the gradient at the centre is 57 km s$^{-1}$ kpc$^{-1}$ compared with an initial value of 41 km s$^{-1}$ kpc$^{-1}$. The increased central concentration of matter is strongly reflected in the significant difference between the rotation curve of our mergers and the progenitor. 

A comparison of central surface brightness between isolated models M1--M5 and mergers M7--M16 shows the merger models exhibiting an elevated surface brightness up to two orders of magnitude greater. L14 showed that BCDs  exhibit surface brightnesses higher than dwarf irregulars by up to three orders of magnitude so our models are consistent with observations. The increase in surface brightness, SFR, and rotation curve gradient throughout our models as well as their morphological resemblance of BCDs clearly allows us to identify that our models enter a BCD phase.

\begin{figure}
\psfig{file=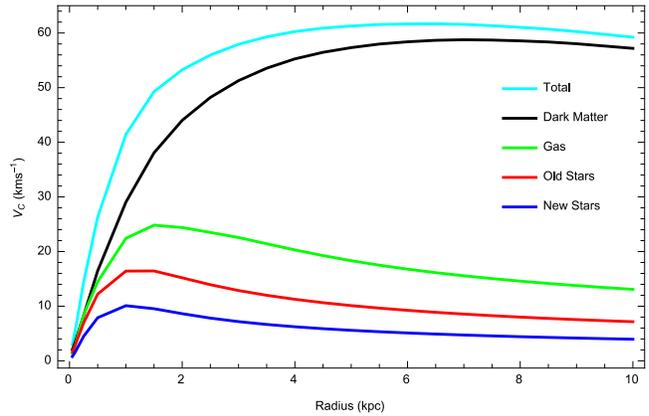,width=8.5cm}
\caption{The final rotation curve of M7 once merging is complete. The gas (purple) contribution shows the greatest change and the contribution from the new stars (blue) is now visible.}
\label{Figure. 4}
\end{figure}

\begin{figure}
\psfig{file=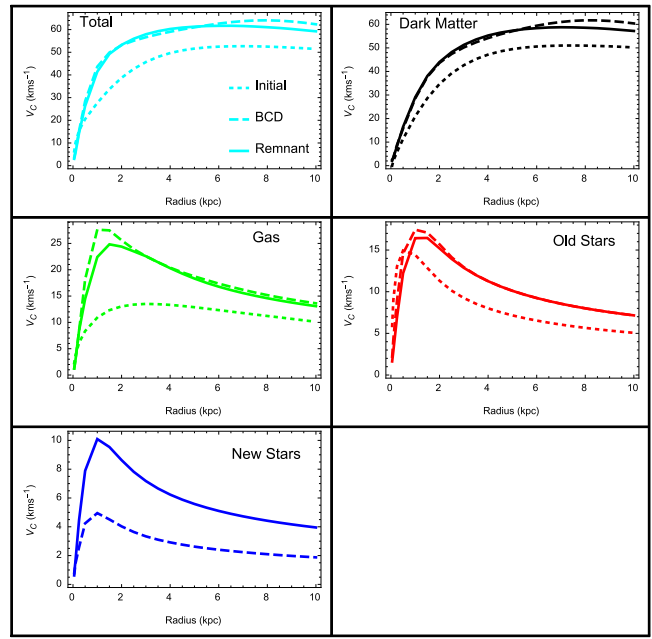,width=8.5cm}
\caption{The evolution of the contributions of each component to the total rotation curve in the initial, BCD and remnant phases in merger model M7. The gas (purple) undergoes significant concentration as reflected in the rotation curve. Dark matter (black) concentration is also visible.}
\label{Figure. 5}
\end{figure}
\subsubsection{BCD model features}
In merger models  M7--M16 we consistently observed increased and steepened rotation curves. We analysed our data at three different time steps for each simulation: the initial at 0 Gyr, the mid simulation at 1.12 Gyr, and the final result at 2.26 Gyr (which from here on we will refer to as the initial, BCD, and remnant phases unless otherwise stated). Fig. 4 is the rotation curve of the remnant of the fiducial model M7 and  can be compared to M1 in the top left-hand panel of Fig. 1. We observe the rotation curve to remain dark matter dominated with a much stronger central contribution from gas, the height increase occurs over the whole curve and gradient increases within 2 kpc from the centre. Fig. 5 shows the evolution of the total curve and the dark and baryonic matter components during each phase. We found the dark matter transitioned from the initial SB profile to a NFW resemblant distribution  and central concentration increased by up to six times the initial value with some models showing up to 0.6 M$_{\odot}$ pc$^{-3}$. This change is reflected in the top right-hand panel of Fig. 5 as the significant increase between the dotted  and dashed curves resembling the initial and BCD phases, respectively. We can see this distribution is maintained into the remnant phase as the solid curve almost matches the dashed. The gas contribution shows the most significant changes to its rotation curve, increasing in both height and gradient at all radii but most notably in the centre where the peak increases by 15 km s$^{-1}$ and the gradient by 20 km s$^{-1}$ kpc$^{-1}$. Both of these are significant percentages of the 22 km s$^{-1}$ and 16 km s$^{-1}$ kpc$^{-1}$ changes observed overall. Clearly this high concentration of gas is responsible for the majority of the increase in rotation curve gradient during the merger.

 \begin{figure}
\psfig{file=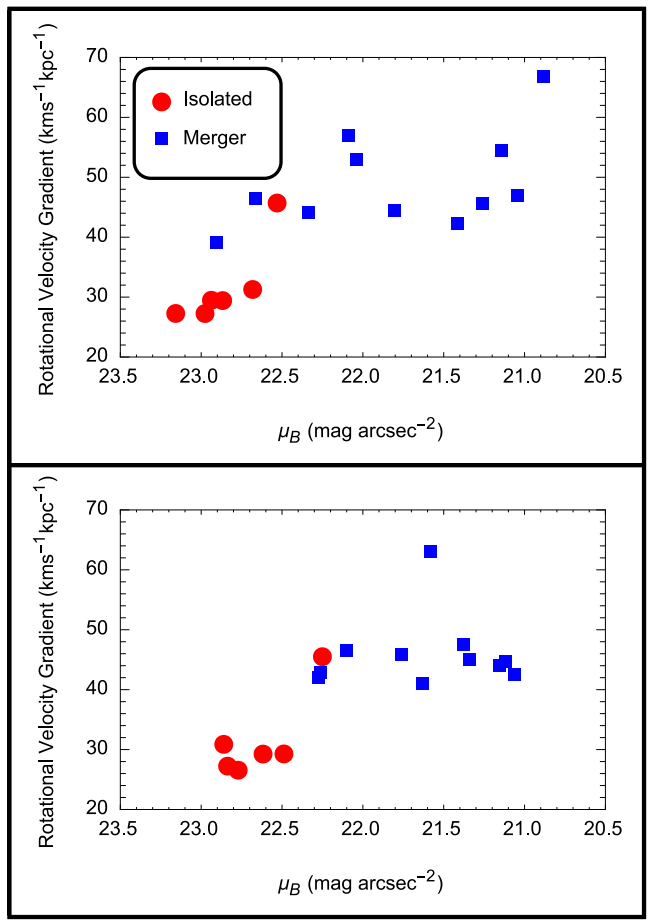,width=8.5cm}
\caption{The relationship between circular velocity gradient and surface brightness in the simulated dwarfs M1 to M16 and M19 using a fixed $M_\odot / L_\odot$. The top panel shows the BCD phase while the bottom shows the remnant. BCDs clearly inhabit a different region to isolated dwarf irregulars as their central surface brightness and rotation curve gradients are much higher. }
\label{Figure. 6}
\end{figure}

 The rotation curve of the old stars shifts and increases its maximum outward slightly to just inside 2 kpc with a slight decrease in gradient as they old stars form an envelope around the core dominated by new stars and gas. After this central concentration, the old stars maintain the same distribution as the initial model as reflected by the matching velocity profile shape. The new stars start off with a low, shallow rotation curve representative of their low numbers due to the only recent starbursts and their peak is within 1 kpc of the centre which correlates with the small optical extent of the star forming region of BCDs. The increase in contribution from new stars to the rotation curve between the BCD and remnant phases matches the decrease observed in the gas contribution implying the star forming region is stable as reflected by the star formation history (SH) in Fig. 3, and is not losing significant amounts of gas due  to feedback. The remnant phase of our models maintains the disturbed morphology and steep rotation curves of the BCD phase. As in Fig. 3 the SFR of some of models M7--M16 remains elevated without significant quenching for Gyr at a time, well into the remnant phase and possibly implies that BCDs can be long lived on Gyr time-scales.
 
  The gas and new star phase metallicity interior to 5 kpc of M7 increases over the course of the simulation with the most rapid increase coinciding with the starburst. Gas phase metallicity drops by $\Delta $log$_{10}[Z_{\rm G}/Z_{\odot}] = 0.1 $ as more pristine gas flows to the interaction centre, igniting the starburst. Metallicity in both M1 and M7 then begins to increase resulting in metallicity differences  $\Delta $log$_{10}[Z_{\rm G}/Z_{\odot}] = 0.075$   and 0.358 during the BCD phase and remnant phases, respectively, with M7 being more metal rich. Comparatively the stellar-phase metallicity for the new stars begins similarly between the two models followed by a rapid increase in metallicity  in M7 compared to M1 with a differences in the BCD phase of $\Delta $log$_{10}[Z_{\rm NS}/Z_{\odot}] = 0.05$ and  0.244 in the remnant phase.  
 
 \begin{figure}
\psfig{file=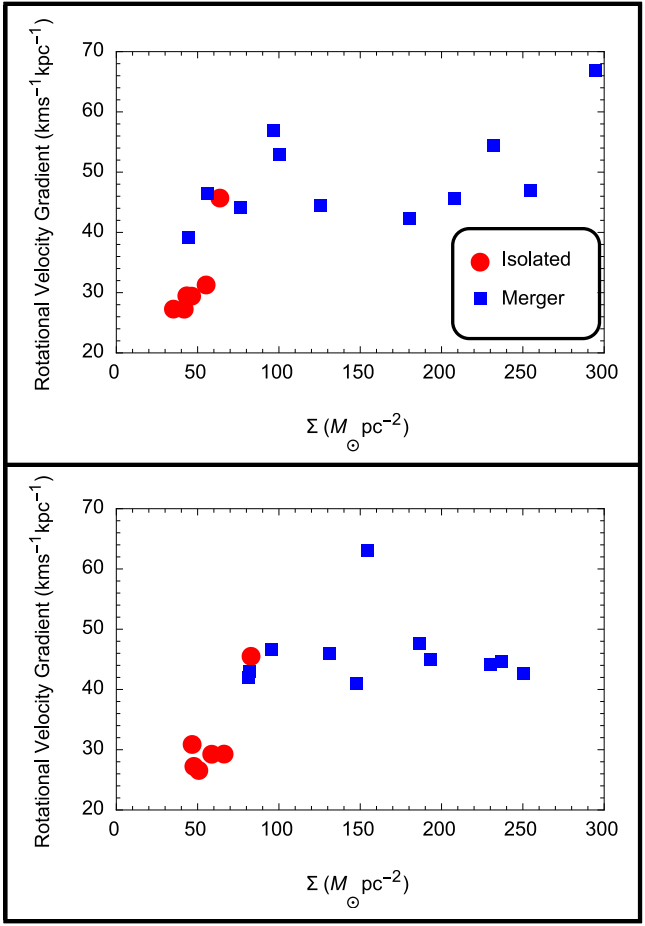,width=8.5cm}
\caption{The same as Fig. 6 but the relationship between circular velocity gradient and projected surface stellar mass density. The BCDs and isolated dwarfs exhibit the same relationship in this figure as Fig. 6. }
\label{Figure. 7}
\end{figure}

 \begin{figure}
\psfig{file=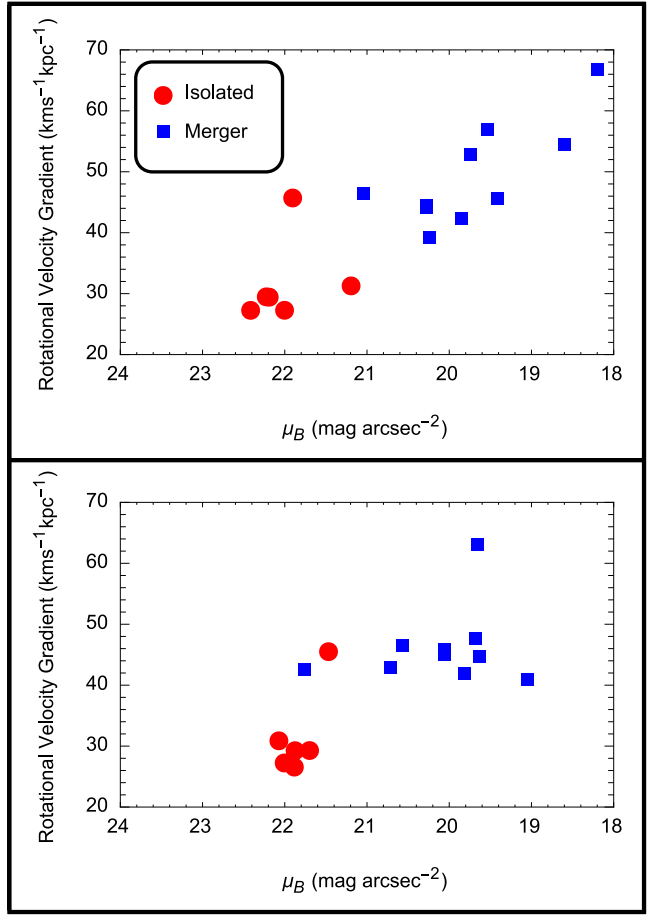,width=8.5cm}
\caption{The same as Fig. 5 but using a varying $M_\odot / L_\odot$. Surface brightness is notably greater and the correlation tighter in the BCD phase (top panel) compared to the constant $M_\odot / L_\odot$. }
\label{Figure. 8}
\end{figure}

\subsection{Correlation between $\mu_{B}$ and rotation velocity gradient}
We can estimate the central \textit{B}-band surface brightness using 
\begin{equation}
\mu_{B} = 27.05 - 2.25{\rm log}_{10}(\Sigma),
\end{equation}
 where we estimate the projected luminosity density ($\Sigma$) by summing the mass of stars within 1 kpc of the centre and assuming a mass to light ratio of    1.  Fig. 6 is the $V_{\rm C}/R_{\rm C}$-$\mu_{B}$ plane for  isolated models M1--M6 and merger models M7--M16 and M19 and was calculated directly from Fig. 7. Fig. 7 shows the plane of circular velocity gradient and projected stellar mass density and we observe the same correlation as Fig .6.  Overall our  $V_{\rm C}/R_{\rm C}$-$\mu_{B}$ plane resembles that of L14's  study where BCDs sit to the top right of dwarf irregulars. Our models show an increased surface brightnesses of up to 2 mag and rotational velocity gradients increased by up to 30  km s$^{-1}$ kpc$^{-1}$. While in the BCD phase models M7--M16 have a rotational gradient varying between 40 and 60 km s$^{-1}$ kpc$^{-1}$ while the remnant phase varies only over 8km s$^{-1}$ kpc$^{-1}$ (this is omitting the initial NFW dark matter profile models).  Between the BCD and remnant phases \textit{B}-band magnitudes become more compact by a maximum of 1 mag arcsec$^{-2}$ while projected stellar mass becomes more compact by 30 M$_\odot$pc$^{-2}$.  We observe a migration towards a common part of the  $V_{\rm C}/R_{\rm C}$-$\mu_{B}$ plane between the BCD and remnant phases as the evolution of the mergers produces a consistently increased SFR and surface brightness over varying time-scales with the common end result of the BCD remnant with a steeper rotation curve  and elevated surface brightness.
The changes observed in the projected stellar mass and surface brightness can be attributed to the elevated SFRs of the BCD phase. Although these values decrease for most models after the BCD phase they decrease gradually and by the end of the simulation are still elevated with respect to the isolated model value. From this we can infer that new stars are continuously being created at an elevated rate allowing surface brightness and projected stellar mass density to remain high. The decrease in rotation curve gradient between phases can be attributed to components settling into more stable orbits. As we produced the points for Figs. 6 and  7 at 1 kpc from the centre any components that have settled to orbits greater than this will no longer be included in the calculation although they will still contribute to the inner rotation curve. Furthermore stars formed from gas  inside the 1 kpc boundary that remain within the bounds will not change the contribution to the rotation curve while adding contribution to surface brightness. Any gas lost from the centre due to feedback will then cause rotation gradient decrease. The gas consistently shows the greatest decrease in gradient contribution between the BCD and remnant phases so the combination of gas conversion to stars, dynamic settling of the system and feedback outflows are a viable cause.

\begin{figure}
\psfig{file=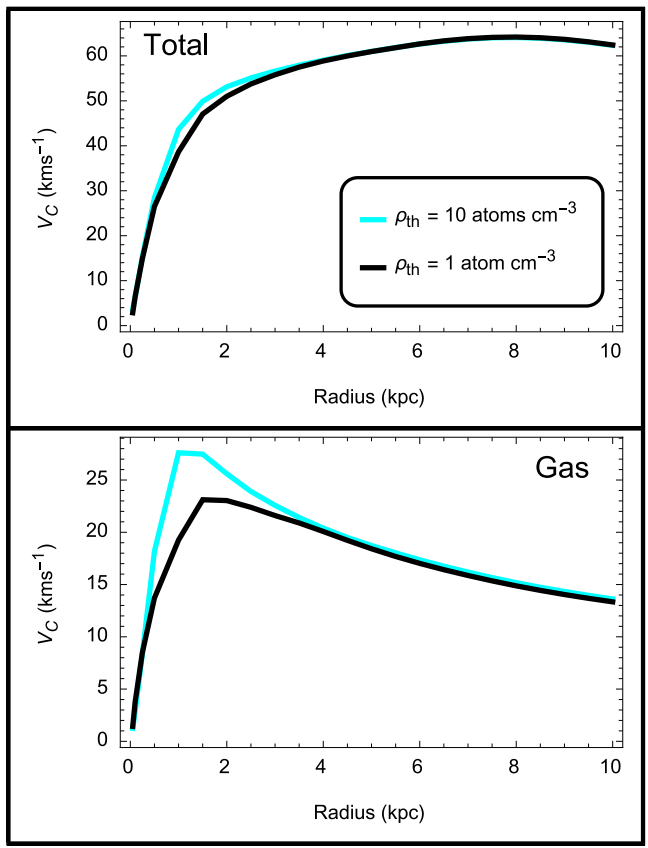,width=8.5cm}
\caption{The effect of star formation threshold density on the total rotation curve (top panel) and gas contribution (bottom panel) for merger models M7 with $\rho_{th} = 10 $ atom cm$^{-3}$ and M8 with with $\rho_{th} = 1 $ atom cm$^{-3}$. We observe that the gas rotation curve is significantly effected. }
\label{Figure. 9}
\end{figure}

\begin{figure}
\psfig{file=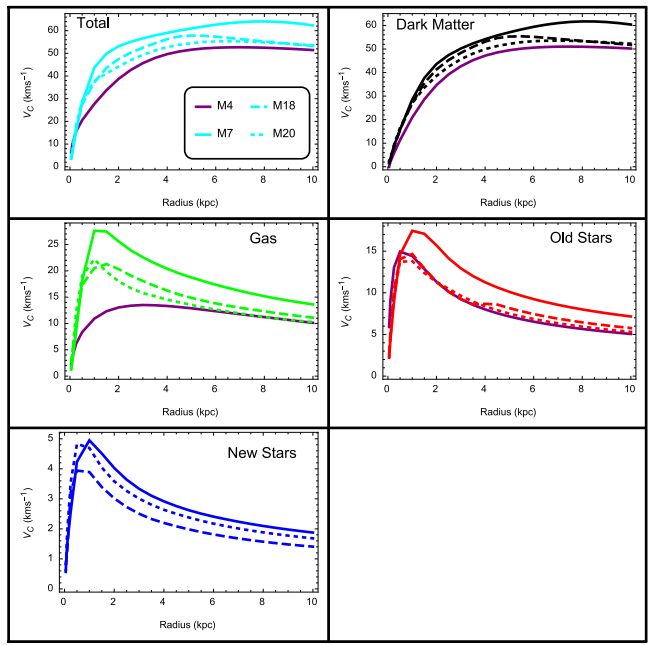,width=8.5cm}
\caption{Comparison of rotation curve contributions of different components in merger models  M7 with m$_{2} = 1$ (major merger), M18 with m$_{2} = 0.3$ (unequal mass merger) and M20 with m$_{2} = 0.1$ (minor merger). The brown curve indicates the initial distribution of each component in the larger mass dwarf and there appears to be a clear correlation between mass ratio and rotation curve height and gradient.}
\label{Figure. 10}
\end{figure}

 We then adopted the age varying \textit{B}-band mass to light ratio from MUISCAT (Vazdekis et al. 2012) using values for metallicity  [Fe/H] $\approx$  [M/H] = --1.3146  and fit a least-squares quadratic regression to produce a continuous $M_\odot / L_\odot$ as a function of time. Fig. 6  was reproduced as Fig. 8  with this new $M_\odot /L_\odot $  for new stars with age $<$1 Gyr and $M_\odot /L_\odot  = 1$ for old stars and new stars with age $>$1 Gyr.  We observe the central surface brightness of the varying $M_\odot /L_\odot $  increases up to 3.62 mag greater compared to isolated models  during the BCD phase  with an average of 2.01 mag across the SB dark matter models. The linear correlation on the $V_{\rm C}/R_{\rm C}$-$\mu_{B}$ plane of the BCDs becomes tighter during the BCD phase and global surface brightness is higher compared to the fixed $M_\odot /L_\odot $. Furthermore all merger models consistently sit to the right of the isolated models as opposed to the constant $M_\odot /L_\odot $ where some merger models exhibited similar surface brightness to some isolated models. Surface brightness decreases in all models between the BCD and remnant phases unlike the fixed $M_\odot /L_\odot $ where some showed increases. 

 Between the BCD and remnant phases the surface brightness distribution becomes more compact by a maximum of  1 mag arcsec$^{-2}$ but now span 1.5 mag arcsec$^{-2}$ more than the remnants of the fixed   $M_\odot /L_\odot $. We observe the same collective evolution to a common part of the $V_{\rm C}/R_{\rm C}$-$\mu_{B}$ plane but the maximum difference in surface brightness between the isolated and merger models is 1 mag arcsec$^{-2}$ greater than the fixed $M_\odot /L_\odot $ and the average remnant surface brightness is 1.5 mag arcsec$^{-2}$ brighter. The spread of surface brightness in the remnant phase increased by 1.5 mag arcsec$^{-2}$.  Velocity gradients remain unaffected and decreases in velocity gradient and surface brightness follow the same arguments presented for dynamical relaxation and slowly decaying SFR.

\subsection{Dependence on Model Parameters}
\subsubsection{Star formation theshold density}
Throughout the models we assigned a star formation threshold density of either 1 or 10 atoms cm$^{-3}$. The top panel of Fig. 9 shows the difference in BCD phase total rotation curve while the bottom panel shows the difference in gas contribution between merger models M7 and M8. The overall rotation curves vary very little outside 3 kpc however M7 is 4 km s$^{-1}$ kpc$^{-1}$ steeper in the centre and rises slightly higher. The gas contribution of M7 is higher by 5 km s$^{-1}$  and steeper by 8 km s$^{-1}$ kpc$^{-1}$, exhibiting a curve that peaks earlier and for longer. Given a high gas threshold density will consume less gas it is not a surprising result that M7 exhibits a steeper curve. Contribution to the total rotation curve from dark matter and old stars vary little between the two models. M7 has a dark matter curve that rises only slightly higher while M8 has an old stars contribution peak of 1 km s$^{-1}$ greater which stays higher for longer. The new stars contribution in M8 is 2 km s$^{-1}$ greater at its peak and remains elevated above M7 out to 10 kpc. This is a result of the difference in star formation threshold and is negligible compared to the variation in gas contribution. Central gradient contribution from the other components vary little between the two models as M8 shows a 1 km s$^{-1}$ kpc$^{-1}$ greater contribution from old and new stars while M7 has a 1 km s$^{-1}$ kpc$^{-1}$ greater contribution from dark matter. It is clear that while the star formation threshold density will affect the rotation curve gradient and contributions from the other components, it doesn't have an affect on the formation of BCDs.

\begin{figure}
\psfig{file=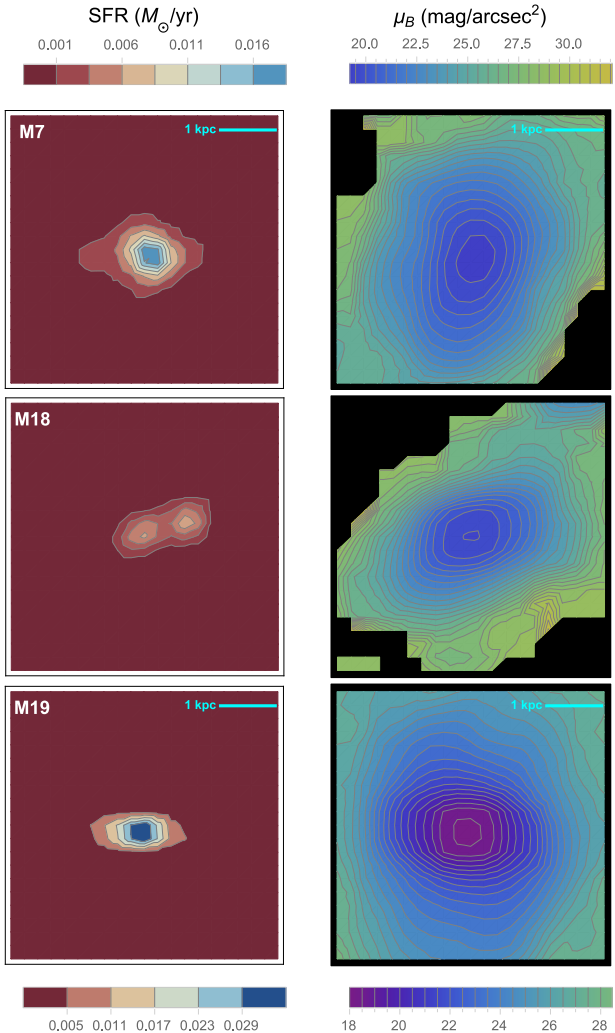,width=8.5cm}
\caption{2D star formation and \textit{B}-band isophotal maps for M7, M18 and M19 centred on the centre of mass of the merger interaction. Each plot is 5 kpc by 5 kpc. The bottom scales apply to the last plots of each column  while the top scales apply to the rest. Note the off centre star formation and surface brightness of the minor merger M18 and the high SFR and $\mu_B$ of the NFW profile M19.  }
\label{Figure. 11}
\end{figure}

\subsubsection{Dark matter profile}
Although most of the models we ran were a SB dark matter merger model,  NFW cusp models were also investigated. The isolated NFW dark matter model M6 can be seen in Fig. 6 and Fig. 7 as the red data point that exhibits an already initially steep central rotation curve. Despite its initially steep rotation curve, the NFW merger M19 entered a BCD phase during the merger and exhibited all the related properties. M19 showed double the gas contribution to the gradient compared to the SB models with the same gas mass fraction and star formation threshold density. The already high dark matter contribution maintained is NFW cusp shape, and both the dark matter gradient contribution and central dark matter density doubled within 2 kpc during the merger. SFR during the BCD phase of M19 was over double that of the SB model M7 shown in Fig. 3 and declined to a similar final value in the remnant phase. The significant increase in central matter density in M19 clearly shows why it has the steepest rotation curve and brightest centre out of all models during the BCD phase. Its central surface brightness dims significantly more between the BCD and remnant phases than the SB models as its higher SFR depletes gas much faster. The remnant phase brightness of M19  is still at the same magnitude as an SB dark matter BCD and the rotation curve gradient remains high implying it is likely still a BCD, but in its very late stages as the SFR and surface brightness are both declining. It is clear that BCD creation is not dependent on the initial morphology of the progenitor dwarfs and the merger process is mainly responsible for the formation of BCDs. The rotation curve does increase more significantly in the NFW model.

\subsubsection{Mass ratio of merging and interactions}
Fig. 10 shows a  comparison of rotation curve contribution for the different components in merger models M7 (mass ration 1), M18 (mass ratio 0.3) and M20 (mass ratio 0.1) at 1.12 Gyr with the initial distribution overlaid. The curves of M18 and M20 are still dark matter dominated, the gas contribution shows central concentration and the old stars distribution stays almost the same. Both models have significantly disturbed morphology at this time with a gas and old stars envelope surrounding irregular regions of new star formation. Although there has not been a significant increase in surface brightness the SFR has tripled in  M18  and quadrupled in M20. The smaller pericentre of M20 allowed it to merge fully by the 1.12 Gyr BCD phase, the SH of this model has the same shape as M7 and the gas distribution exhibits some tidal features. M18 at this time has not fully merged, the smaller companion's core of mainly old with some new stars is currently just outside the core of the larger dwarf and can be seen as the small bump around 3.5 kpc in its old stars rotation curve. The SH of this model quickly dips after a spike around 1.12 Gyr, and then steadily increases to the same value at the spike as the merger completes and builds again to another BCD resemblant phase over the span of a Gyr.

Looking the top left panel of Fig. 10 we can clearly see the effect of difference in mass ratio on the rotation curves of our BCDs. Total curves are lower and shallower with smaller the mass ratio and gas contributions in the middle left panel still contribute the majority of the change in rotation curve gradient, but the gas in M18 is much more spread out due to incomplete merging at this time. The old stars show the same trend with mass ratio due to a diminishing number present and a longer merger time as the mass ratio decreases. Dark matter concentration is significantly weaker in these models and we can see this in the relatively similar total rotation curves M18 and M20 sitting well below the curve of M7. We find that the distribution of dark matter in both cases exhibits the same change from the initial SB to a NFW resemblant cusp. The dark matter concentration of M18 exhibits a plateau from 1 to 3  kpc after its high central value. Central dark matter density increases by a factor of 2 in M18 as we have a longer, weaker interaction and the merging is not complete when we reach the BCD phase at t = 1.12 Gyr.  M20 shows a factor of 4 increase as the small pericentre merge allows for a stronger dark matter concentration.  The interaction models we ran did not result in significantly elevated SFRs nor did they concentrate matter in the progenitor(s). Instead what resulted was either a destroyed system or a dwarf where significant amounts of gas had been scattered,  and neither of these scenarios resulted in heightened or steepened rotation curves. We conclude that interaction models will not produce BCDs as suggested by B08 and will not talk about them again in this paper.

\subsection{Surface brightness and star formation rate}
 Fig. 11 shows the star formation rate within the last 30 Myr during the BCD phase and \textit{B}-band surface brightness both distributed over a 5 kpc by 5 kpc area centred on the centre of mass of fiducial merger M7, minor merger M18 and NFW dark matter profile  merger M19. We observe clear spatial correlations between elevated surface brightness and the regions of the highest star formation rate. From Fig. 2 we note that irregular envelope of old stars is approximately 5 kpc in diameter which would fill an entire panel of Fig. 11 demonstrating that the most active star formation is extremely concentrated compared to the size of the host merger while lower level star formation is ignited over approximately half the BCD. For equal mass mergers we consistently observe central starbursts but M18 shows a more asymmetric,  clumpy distribution of star formation due to the relative weakness of the minor merger interaction and incomplete merging as mentioned in 3.3.3. Part of the core of the minor galaxy can be seen in the top left of the middle right panel of Fig. 11 as a region of slightly higher surface brightness but inert of star formation (middle left panel) as it has been stripped of all gas. 

 The surface brightness of our BCDs are irregular in shape but overall uniform in their central intensities. The BCDs of K14 exhibit small areas of specifically elevated brightness distributed in a larger, irregular region of increased brightness. These brightness increases in K14 are distributed both centrally and irregularly and we can identify the general shapes of our SFR/$\mu_B$ maps with the B-V colour maps of K14. M19 could be related to  UM323 with a bright centre and extended lower level brightness elevation present throughout both galaxies. The less intense but still elevated central brightness of  M18 and Mk900 could be identified although M18 lacks the off-centre brightness 'knot' that Mk900 exhibits, but M18 exhibits this in its SFR map. The oval shaped central brightness of both Mk324 and M7 could also be identified as they are of similar shape and make up a similar amount of their host galaxy.

\section{Discussion}

\subsection{The role of dwarf--dwarf merging in the formation
of steep rotation curves of BCDs}

Observational studies of BCD galaxies show steeply rising rotation curves (e.g. L14) that have been modelled best with dark matter cores (M98; Elson et al. 2010) that dominate the rotation curve.
However it is not explicitly clear how they are formed and what the origin of their steep rotation curve is. 
As discussed in the previous section,
our simulations of dwarf galaxy mergers have produced BCDs with strong resemblance to observational data:
H{\sc i} maps of BCDs (e.g. M98) show a strong central concentration of gas.
Dwarf mergers have been shown to effectively concentrate gas (B08, B15)
and the formation of BCDs from our models through dwarf galaxy mergers with steep central gas rotation curves provides solid support for this process as a candidate for the progenitors of BCDs. Furthermore all models exhibit the same dark matter rotation curve dominance
in their outer parts ($R>2$ kpc)
that provides the majority of the curve shape while it is the concentration of gas during the merger that contributes to the majority of the increase in rotation curve gradient between the initial and BCD phases.

The minor merger models M18 and M20 show different dark matter concentrations that appear to depend on their interaction rather than mass ratio. M20 had a faster merge time-scale due to its small interaction pericentre and  showed a dark matter concentration increase similar to some of the major mergers.  M18 had not fully merged and therefore showed a comparatively smaller concentration. Despite this we still observed dark matter dominated rotation curves where gas concentration provided almost all of the gradient increase in the total rotation curve. The morphology of these models was significantly more disturbed than the major models and they sit near the isolated models on the $V_{\rm C}/R_{\rm C}$-$\mu_{B}$ plane. L14 showed that BCDs with off centred starbursts are placed in the same area as dwarf irregulars on this plane so our results may suggest that we can form off centre starburst BCDs from minor mergers, however as we have only a few of these models further study will be required.

\subsection{dwarf--dwarf merging as a mechanism for BCD formation}

The merger rate of dwarf galaxies in the local Universe is a still mostly unexplored topic. The merger rate of satellite  dwarf galaxies  with $10^{6} _\odot<M<10^{9} _\odot$ within the virial radius of host galaxy should be approximately 10\% in the Local Group and 15\%-20\% in field galaxies with $M>10^{6} _\odot$  since z=1, implying that dwarf mergers are more common further away from massive hosts   (Deason et al. 2014). Without a highly complete survey of dwarf galaxies in both clusters, field and in isolation combined with a well known dwarf merger rate in the different environments it is difficult to comment on the fraction of BCDs produced by dwarf--dwarf mergers. Surveys of dwarf galaxies put the total BCD fraction at approx. 17 \% over a wide range of environments (Hunter \& Elmegreen 2004, 2008) and Drinkwater et al. (1994) put the BCD fraction of the Virgo cluster at around 8\%. Meyer et al. (2014) found cluster BCDs should not exhibit transient starburst behaviour so our results of longer time-scale BCD lifetime agrees with this result somewhat. If the dwarf merger rate in clusters decreases to the present day 8\% is not an unreasonable number as dwarf galaxies would have been `consumed' faster early on.  This introduces the possibility for multiple mergers and the redder LSB population observed in Virgo BCDs by Meyer et al. (2014) potentially supports this. Furthermore Sabatini et al. (2005) found dwarf galaxies on the outskirts of clusters showed more variation in colour compared to their more central counterparts so galaxy location is also likely a factor. Conversely stripping mechanisms can also result in a redder dwarf population (Sabatini et al. 2005) so speculation is the most we are able to do at this stage.

\subsection{Does star formation influence rotation curve profiles?}

The evolution of our rotation curves  between the BCD and remnant phases shows very little evolution regardless of whether the SFR remained high or decayed. Furthermore we observe little evolution on the $V_{\rm C}/R_{\rm C}$-$\mu_{B}$ plane implying minor redistribution of mass  when considering both fixed and variable $M_\odot /L_\odot  $. Lelli et al. (2012b) concluded that unless there is significant mass redistribution  the descendants of BCDs would have to be compact dwarf irregulars. Given that compact dwarf irregulars populate a similar region to BCDs on L14's plot of the $V_{\rm C}/R_{\rm C}$-$\mu_{B}$ plane,  the SH of many of our models declines after the BCD phase, and all other properties remain resemblant  of BCDs, we could also reach this conclusion. 

Almedia et al. (2008) suggests there could be a  distinction between the low level continuous star formation in quiescent BCDs and the periods of episodic starbursts during active phases.  We observe the gas contribution to the rotation curve is significantly greater in models with a high star formation threshold density. This dependence could imply that a newly formed BCD is likely to have a much steeper rotation curve and kinematically show more signs of merging. Conversely we would expect a BCD that has evolved through numerous starburst and quiescent phases to have a shallower rotation curve and a more regular structure due to dynamic settling and gas consumption.

\subsection{Differences in dark matter distributions
between BCDs and other dwarfs}
Several previous studies (e.g.,
Meurer et al. 1996;  M98; Elson et al. 2010)
modelled BCDs and found central dark matter densities up to 10 times greater than isolated dwarfs. 
Our merger models M7 to M16 and M19 show up to a 6 x increase in dark matter concentration compared to M1 to M6 and this is what provides the base rotation curve and gradient for the BCD. While these mass models of BCDs typically show a central concentration of 0.1 M$_{\odot}$pc$^{-3}$ our SB models M7 to M16 show up to 6 times this value whereas M1 to M5 show initial concentrations comparable to these observed densities. With dark matter concentration being initially high it is understandable that our BCDs exhibit an even higher density from the concentration during the merger process and the concentration difference does reflect observations to some extent, however the numerical value of the density does not. The failure of the isolated models to form BCDs suggests that the numerical value of the dark matter concentration isn't as important in forming a BCD but rather the redistribution and concentration of matter is required, a similar conclusion suggested by L14. Our NFW model maintained its cusp shape during the BCD phase and the central density doubled. Whether it is the initially high central density of dark matter limiting the concentration or another phenomena is not possible to say as we only have one NFW model. All the features resemblant of BCDs were still observed so it appears that dark mater distribution does not affect the formation of BCDs, rather it will contribute to the kinematics and surface brightness.

Dark matter concentration has been observationally correlated with starburst regions (e.g. McQuinn et al. 2015) and the concentration of dark matter, central formation of new stars and elevated SFR supports this suggestion of dark matter influenced star formation. While the current understanding of BCD starburst lifetime is that it should not last more than 10 Myr (e.g. Thornley et al. 2000), some of our models exhibit star formation histories that  appear to show BCD phases on Gyr time-scales such as M7 in Fig. 3, M19,   and the episodic SH of M18 that builds again over a Gyr.  The work of V14 shows that BCDs can be produced by a variety of in-falling gas clouds and ignited starbursts on 100 Myr time-scales from the interaction with the potential for longer time-scale star formation caused by supernovae outflows and shocks. V14 produced some BCD galaxies with Gyr starbursts where the stellar metallicity remained approximately the same as the stellar metallicity of the host galaxy for the duration of the burst. The BCDs of K14 showed elevated metallicity in off-centre star forming regions and K14 implied stable, self enriching star formation was being caused by the presence of dark matter.  

Our models M7 to M16 and M19 show centrally correlated starbursts and dark matter dominated centre of mass however  the off-centre star formation morphology was exhibited by  our minor mergers M18 and M20. While we observed dark matter concentration in M18 and M20, the central dark matter concentration increased by a factor of 4 at most. M18 exhibited  irregularly shaped star forming regions with a dark matter concentration plateau out to 3 kpc  but an episodic SH implies the irregularly distributed star formation is only Myr in age. The observations of strong concentration of dark matter in models M7 to M16 and M19 correlated with spatially compact, elevated SFR considered along with the continually increasing stellar phase metallicity of the new stars formed in the BCD  potentially provides support for K14's suggestions of dark matter influenced stability  in self enriching long lived starburst regions. As we only have a very small model set of minor mergers we cannot comment significantly on the finer details  of lower mass ratio interactions as there appears to be a dependence on the interaction pericentre and mass ratio.

\section{Conclusions}
We have investigated the rotation curves of BCDs formed from 
mergers between gas-rich dwarf irregular galaxies through numerical 
simulations.  Using 20 different dwarf--dwarf merger models of BCD formation,
we have
investigated the correlation between rotation curve gradient and surface brightness, dark matter and  gas concentration, the dependence of rotation curve on star formation threshold and merger mass ratio. The principal results
are as follows:

(1) The slowly rising rotation curves of initial dwarf irregulars
can be transformed into steeply rising ones in the formation of BCDs.
The rotation curve gradients of BCDs with strong central 
starbursts can be by 20 km s$^{-1}$
steeper than those of initial dwarfs.
This result implies that dissipative dwarf--dwarf merging
can be responsible for the observed steep rotation curves of BCDs.

(2) Although the rotation curves in
the outer parts ($>2$ kpc) of BCDs are dominated by dark matter,
both baryonic and dark matter contribute equally
to the rotation curves in the central regions ($R<1$ kpc).  
Dissipative merging process can enhance the central mass
concentration of dark matter by a factor of 6,
and consequently transform dark matter haloes with initially cored distributions
into cuspy ones that are  similar to the NFW profile.

(3) Dark matter provides the base rotation curve 
but the increase in gradient between the progenitor and 
the BCD is due to the strong central concentration of  gas. 
This result suggests that baryonic physics is a key for better
understanding the origin of the rotation curves of BCDs.

(4) BCD rotation curves are dependent on star formation threshold 
density and merger mass ratio. 
BCDs formed from major dwarf--dwarf merging can show steeper rotation
curves that minor mergers.
Simple tidal interaction without merging 
cannot create BCDs with steep rotation curves,
minor mergers are more likely to create BCDs with off centre starbursts.

(5) BCDs formed from dwarf--dwarf merging
are likely to evolve into compact 
dwarf irregulars with a significant amount of residual star
formation  after strong  starburst phases, because there is still a plenty
of gas left in the merger remnants.
The BCDs inevitably  show both high surface brightness 
and steep rotation curves, because efficient gas infall into the 
central regions of merging can drive both central mass concentration
and enhanced star formation.

\section{Acknowledgements}
A. Watts and K. Bekki are grateful to the anonymous referee(s) for their constructive and useful comments that improved this paper. 

\label{lastpage}

\end{document}